\documentclass[12pt,a4paper,reqno]{article}
\usepackage{amsmath}
\usepackage{amssymb}
\usepackage{bbm}
\usepackage{cite}
\usepackage{hyperref}
\hypersetup{colorlinks,linkcolor=blue,citecolor=red,urlcolor=blue}

\allowdisplaybreaks
\numberwithin{equation}{section}

\newcommand{\+}{\mskip2mu} % 1mu
\newcommand{\tab}{\quad\,}

\newcommand{\e}{\mathrm{e}}
\newcommand{\I}{\mathrm{i}}

\newcommand{\cA}{\mathcal{A}}
\newcommand{\cF}{\mathcal{F}}
\newcommand{\cK}{\mathcal{K}}
\newcommand{\cL}{\mathcal{L}}
\newcommand{\cM}{\mathcal{M}}
\newcommand{\cN}{\mathcal{N}}
\newcommand{\cT}{\mathcal{T}}

\newcommand{\SO}{\mathit{SO}}
\newcommand{\SU}{\mathit{SU}}
\newcommand{\Sp}{\mathit{Sp}}

\newcommand{\U}{\mathit{U}}

\newcommand{\ep}{\varepsilon}

\newcommand{\si}{\sigma}

\newcommand{\bb}{\bar{b}}
\newcommand{\bz}{\bar{z}}
\newcommand{\txi}{\tilde{\xi}}
\newcommand{\p}{\partial}

\newcommand{\lp}{\begin{pmatrix}}
\newcommand{\rp}{\end{pmatrix}}
\newcommand{\frc}[2]{\frac{\raisebox{-2pt}{$#1$}}{#2}}
\newcommand{\arXiv}[1]{\href{http://arxiv.org/abs/#1}{arXiv:#1}}

\DeclareMathOperator{\re}{\mathrm{Re}}
\DeclareMathOperator{\im}{\mathrm{Im}}
\DeclareMathOperator{\tr}{\mathrm{tr}}

\newcommand{\beq}{\begin{equation}\begin{aligned}}
\newcommand{\eeq}{\end{aligned}\end{equation}}
\newcommand{\bi}{\begin{itemize}}
\newcommand{\ei}{\end{itemize}}
\newcommand{\bea}{\begin{eqnarray}}
\newcommand{\eea}{\end{eqnarray}}
\newcommand{\ba}{\begin{array}}
\newcommand{\ea}{\end{array}}
\newcommand{\bt}{\begin{tabular}}
\newcommand{\et}{\end{tabular}}
\newcommand{\bc}{\begin{center}}
\newcommand{\ec}{\end{center}}

\newcommand{\cG}{\mathcal{G}}

\newcommand{\CF}{\hat\cF}
\newcommand{\cref}{{\bf [check ref]}}

\def\nv{{n_{\rm v}}}
\def\nh{{n_{\rm h}}}

\def\stt {$\SU(3) \times \SU(3)$}

\newcommand{\M}{M}

\newcommand{\dd}{\mathrm{d}}
\newcommand{\ee}{\mathrm{e}}
\newcommand{\ii}{\mathrm{i}}

\begin{document} %%%%%%%%%%%%%%%%%%%%%%%%%%%%%%%%%%%%%%%%%%%%%%%%%%%%%%%

\begin{titlepage}

\rightline{\small IPhT-T14/016}
\rightline{\small ZMP-HH/14-08}
\rightline{\small ITP-UH-06/14}
\rightline{\small Imperial/TP/2014/DW/02}
\vskip 0.5cm

\begin{center}
 {\large\bfseries Quantum Corrections in String Compactifications on \\[.5ex] SU(3) Structure Geometries} \\[3ex]
 \textbf{Mariana Gra\~na$^a$, Jan Louis$^{b,c}$, Ulrich Theis$^d$, Daniel Waldram$^{e}$} \\[2ex]
 \slshape
 ${}^a$Institut de Physique Th\'eorique, CEA/Saclay \\ 91191 Gif-sur-Yvette Cedex, France \\ \href{mailto:mariana.grana@cea.fr}{mariana.grana@cea.fr} \\[2ex]
 $^{b}$Fachbereich Physik der Universit\"at Hamburg \\ Luruper Chaussee 149, 22761 Hamburg, Germany \\[.5ex]
 {}$^{c}$Zentrum f\"ur Mathematische Physik, Universit\"at Hamburg \\ Bundesstrasse 55, 20146 Hamburg, Germany \\
 \href{mailto:jan.louis@desy.de}{jan.louis@desy.de} \\[2ex]
 ${}^d$Riemann Center for Geometry and Physics, Leibniz Universit\"at Hannover, \\ Appelstrasse~2, 30167 Hannover, Germany \\
 \href{mailto:ulrich.theis@riemann.uni-hannover.de}%
 {ulrich.theis@riemann.uni-hannover.de} \\[2ex]
 $^{e}$Department of Physics, Imperial College London \\ London, SW7
 2AZ, U.K. \\ 
\href{mailto:d.waldram@imperial.ac.uk}{d.waldram@imperial.ac.uk}
\end{center}

%\vskip 1cm
%\medskip

\begin{center} {\bf ABSTRACT } \end{center}
\vspace{-2mm}

\noindent
We investigate quantum corrections to the classical
four-dimensional low-energy effective action of type II string theory
compactified on $SU(3)$ structure geometries. Various methods
previously developed for Calabi--Yau compactifications are adopted to
%determine 
constrain~--~under some simple assumptions about the low-energy degrees
of freedom~--~the leading perturbative corrections to the moduli space
metrics in both $\alpha'$ and the string coupling constant. 
% We find -- in complete analogy to the Calabi--Yau case -- that the
% corrections take a universal form dependent only on the Euler
% characteristic of the six-dimensional compact space. 
%
We find that they can be parametrized by a moduli dependent function
in the hypermultiplet sector and a constant in the vector multiplet sector.
We argue that 
under specific additional assumption they take~--~in 
complete analogy to the Calabi--Yau case~--~a 
universal form which depends only on the Euler
characteristic of the six-dimensional compact space. 

\vfill

June 2014

\end{titlepage}

\section{Introduction} %%%%%%%%%%%%%%%%%%%%%%%%%%%%%%%%%%%%%%%%%%%%%%%%%

In this paper we consider type II string theory in backgrounds
$M_{1,3}\times X_6$, where $M_{1,3}$ is a  four-dimensional ($d=4$)
Lorentzian space-time and $X_6$ is a compact
six-dimensional manifold. Demanding that the effective theory in
$M_{1,3}$ admit eight unbroken supercharges ($\mathcal{N}=2$ in $d=4$)  
constrains $X_6$ to be a manifold with $\SU(3)$
structure~\cite{waldram,GLMW}.\footnote{\label{foot:stt}More generally, viewed as generalized geometries~\cite{Hitchin,Chris,EGG},  $\mathcal{N}=2$ supersymmetry requires an \stt\
structure on the generalized tangent space~\cite{GMPT2,JW,GLW1,GLW2} or even more generally an $\SU(6)$
structure on the exceptional generalized tangent space~\cite{GLSW}.}
%(We review some of their properties in section~\ref{stt}.)  
Manifolds with $\SU(3)$ structure admit a globally defined
nowhere-vanishing spinor and as a consequence the structure group of
the tangent space is reduced, meaning that it can be patched using
only an $\SU(3)$ subgroup of $\SO(6)$. If in addition the spinor is
covariantly constant the Levi--Civita connection has $\SU(3)$ holonomy,
and in this case $X_6$ is a Calabi--Yau manifold. 

The $\mathcal{N}=2$ low-energy effective Lagrangian of such
backgrounds has been computed in
refs.\ \cite{Louis:2002ny,GLMW,GLW1,GLW2,GLSW,KPM,DAuria,MK,BC,CKP} 
at the string tree level and for ``large'' manifolds~$X_6$ assuming a
suitable Kaluza--Klein reduction. For Calabi--Yau manifolds this is
straightforward in that the massless modes are in one-to-one
correspondence with the cohomology of $X_6$, and the resulting
$\mathcal{N}=2$ supergravity is ungauged with no potential~\cite{CdO}.
In the generalized case the corresponding analysis 
% of the Kaluza--Klein
% spectrum and the computation of the effective action 
is much harder and can only be performed 
if there is a suitable hierarchy of the low-energy modes.
In this case the  effective action is a gauged  $\mathcal{N}=2$
supergravity with a potential which stabilizes (some of) the
moduli. 
%However, the existence of a supersymmetric vacuum is by no
%means guaranteed. 

The low-energy effective Lagrangian is corrected by higher-order
$\alpha'$ terms as well as by string loops ($g_s$ corrections).
Both types of corrections can be parametrized by a
scalar field (or a modulus) in the low-energy effective action.
Higher-order $\alpha'$ corrections arise when the size of the manifold is
comparable to the string length and thus they are controlled by the
volume modulus of $X_6$. String loop corrections on the other hand 
are counted by $g_s$ with the dilaton being the corresponding
modulus. For Calabi--Yau compactifications both types of corrections
have been partially computed, and we review some of these results in the
following. However, for generalized compactifications little is known
about these corrections primarily due to the fact that 
%such compactifications are generically off-shell and so 
they have no direct
worldsheet description.  

The purpose of this paper is to investigate what can be said about 
$\alpha'$ and $g_s$ corrections in the specific case of $\SU(3)$
structure compactifications. We will not focus on a particular
background but rather attempt to constrain the generic form of the
possible corrections, under some simple assumptions. We also do not
consider any non-trivial NSNS or RR fluxes, or more general \stt\
structures (see footnote \ref{foot:stt}). 

Using simple extensions of the
arguments employed for Calabi-Yau compactifications in \cite{Strominger,Antoniadis:1997eg,Gunther:1998sc,Becker:2002nn,AMTV,loop}
we find -- given there is an expansion in terms of a finite set of
low-energy fields -- that the $\cN=2$ supersymmetry strongly
constrains the leading corrections to the kinetic terms.
In the hypermultiplet sector a non-renormalization theorem holds
and the metric is perturbatively  corrected only at one-loop by
a moduli dependent function. In the type IIA 
vector multiplet sector we find $\alpha'$ corrections 
which in a large volume expansion correct the holomorphic
prepotential by a constant term at leading order. 
In type IIB the vector multiplet sector on the other hand 
is not corrected at all.
With the additional assumption of a putative mirror symmetry
we are able to argue that all corrections are proportional to the 
Euler characteristic of
$X_6$ exactly as in Calabi-Yau compactifications.

Generically, $\SU(3)$ structure compactifications admit fewer
Peccei--Quinn (PQ) symmetries than Calabi--Yau compactifications,
since some of the corresponding moduli acquire a mass from the gauging
of the $\mathcal{N}=2$ supergravity. Remarkably, we find that the
leading kinetic energy corrections still have the full set of PQ
symmetries, even under shifts of the massive moduli.   

The paper is organized as follows: In section~2 we review some basic
facts about $\mathcal{N}=2$ supergravity and $\SU(3)$ structure
compactifications. We also summarize our assumptions about the moduli
space of low-energy fields. Section~3 then discusses the form of
the leading corrections to the metrics on these moduli spaces: the
prepotential for the vector multiplet fields, and the
quaternionic-K\"ahler metric for the hypermultiplet fields. We first
discuss the constraints on the metric of the geometric moduli that
arise from dimensionally reducing the known ten-dimensional $R^4$
correction terms. We then
turn to the full vector and hypermultiplet moduli space metrics, first
for the $\alpha'$ corrections and then for the $g_s$
corrections. Section~4 summarizes our results and 
contains some concluding remarks.

\section{Preliminaries}\label{prelim}

\subsection{$\mathcal{N}=2$ supersymmetry in $M_{1,3}$}\label{N=2}

In order to discuss the $\alpha'$ and $g_s$ corrections we need to
briefly assemble some facts about $\mathcal{N}=2$ supergravity (for a
review see e.g.\ \cite{Andrianopoli:1996cm}). A generic spectrum
contains a gravitational multiplet, $\nv$ vector multiplets and $\nh$
hypermultiplets.\footnote{Tensor multiplets are dual to hypermultiplets if they are massless and to vector multiplets if they are massive.} 
The gravitational multiplet 
%$(g_{\mu\nu},\Psi_{\mu {\cal A}}, A_\mu^0)$ 
contains the spacetime metric,
%$ g_{\mu\nu}, \mu,\nu =0,\ldots,3$, 
two gravitini 
%$\Psi_{\mu {\cal A}}, {\cal A}=1,2$, 
and the graviphoton $A_\mu^0$. A vector multiplet 
%$(A_\mu,\lambda^{\cal A}, t)$ 
contains a vector $A_\mu$, two gaugini 
%$\lambda^{\cal A}$  
and a complex scalar $t$. Finally, a hypermultiplet 
%$(\zeta_{\alpha}, q^u)$ 
contains two hyperini 
%$\zeta_{\alpha}$ 
and four real scalars $(q^1,q^2,q^3,q^4)$. 

The $\mathcal{N}=2$ supersymmetry enforces the scalar field space $\M$ 
locally to be a product
\beq\label{VHsplit}
\M= \M_{\rm v}\times \M_{\rm h} \ ,
\eeq
where $\M_{\rm v}$ is a $2\nv$-dimensional manifold spanned by the
complex scalars $t^i, i= 1,\ldots,\nv$, contained in $\nv$ vector
multiplets, while $M_{\rm h}$ is a $4\nh$-dimensional manifold spanned
by the real scalars $q^u, u= 1,\ldots,4\nh$, in $\nh$
hypermultiplets. Thus their sigma-model Lagrangian is of the form 
\begin{equation}\begin{aligned}\label{sigmaint}
{\cal L}\ =\  
%- \mathrm{i} \mathcal{N}_{IJ}\,F^{I +}_{\mu\nu}F^{\mu\nu\, J+}
%+ \mathrm{i} \overline{\mathcal{N}}_{IJ}\,
%F^{I-}_{\mu\nu} F^{\mu\nu\, J-}
g_{i\bar \jmath}(t,\bar t)\, D_\mu t^i D^\mu\bar t^{\bar \jmath}
+ h_{uv}(q)\, D_\mu q^u D^\mu q^v
- V(t,\bar t,q) \ ,
\qquad \mu=0,\ldots,3\ ,
\end{aligned}\end{equation}
where  $g_{i\bar \jmath}$ is the metric on $\M_{\rm v}$ while $h_{uv}$
is the  metric on $\M_{\rm h}$. In general, isometries on
$\M_{\rm v}$ and $\M_{\rm h}$ can be gauged, so that $D$ is an
appropriate covariant derivative which includes the couplings of the
charged scalars to the vector bosons. Only if the theory is gauged
can it admit a non-trivial potential $V$, which furthermore is
completely determined by the choice of gauging.

Supersymmetry further constrains $\M_{\rm v}$ to be a
special-K\"ahler manifold, so that the metric can be written
as~\cite{deWit:1984pk,Craps:1997gp}  
\begin{equation}\label{gdef}
g_{i\bar \jmath} = \partial_i \partial_{\bar \jmath} K^{\rm v}
\qquad \text{with}\qquad
K^{\rm v}= -\ln \I\left( \bar X^I \cF_I - X^I\bar \cF_I \right)\ .
\end{equation}
Both $X^I(t)$ and ${\cal F}_I(t)$, $I= 0,1,\ldots,\nv$, are
holomorphic functions of the scalars $t^i$, and 
%in the ungauged case one can always choose
$\cF_I = \partial\cF/\partial{X^I}$ is the derivative of a holomorphic prepotential $\cF(X)$ homogeneous of degree two. Furthermore, it is possible to adopt a system of `special coordinates' in which $X^I= (\I,t^i)$ (see e.g. \cite{Craps:1997gp} for further details).

$\M_{\rm h}$ is similarly constrained to be a quaternionic-K\"ahler
manifold, which means its holonomy group is of the form $\Sp(1)\times
H$ with $H\subset\Sp(\nh)$~\cite{Bagger:1983tt,deWit:1984px}.   
There is a special class of quaternionic-K\"ahler manifolds which
arise in string tree-level effective actions known as `special
quaternionic-K\"ahler manifolds' $M_{\rm SQK}^{4\nh}$. These manifolds 
can be viewed as a $2\nh$-dimensional torus fibred over 
a special-K\"ahler base, of the form 
$[\SU(1,1)/\U(1)]\times M_{\rm SK}^{2\nh-2}$, where 
$M_{\rm SK}^{2\nh-2}$ is a special-K\"ahler manifold of dimension
$2\nh-2$. This relation between special-K\"ahler and quaternionic
K\"ahler manifolds is known as the `c-map'~\cite{CFG}
 \begin{equation} \label{c-map}
  \frc{\SU(1,1)}{\U(1)}\ \times\ M_{\rm SK}^{2\nh-2}\ 
  \mapsto\ M_{\rm SQK}^{4\nh}\ .
 \end{equation}
The metric on $M_{\rm SQK}^{4n}$ takes an explicit form known as the
Ferrara--Sabharwal metric, which reads \cite{FS}
\begin{equation}
\begin{aligned}\label{FS}
   \cL & = 
      (\partial_\mu\phi)^2 + \e^{4\phi} \big(\partial_\mu\si - \xi^I \partial_\mu\txi_I \big)^2 
      + g_{a\bb}\, \partial_\mu z^a \partial^\mu\bz^{\bb}  \\
   & \qquad 
      - \e^{2\phi} \big(\cN + \bar{\cN} \big)^{IJ}
         \big( \partial_\mu\txi_I - 2\I\+ \cN_{IK}\+ \partial_\mu\xi^K \big)
         \big( \partial^\mu\txi_J + 2\I\+ \bar{\cN}_{JL}\+ \partial^\mu\xi^L \big)\ ,
\end{aligned}
\end{equation}
where $\phi$ and $\si$ span the $\SU(1,1)/\U(1)$ factor
of~\eqref{c-map}, $\xi^I$ and $\txi_I$ are real coordinates of the
torus fibre, and  $z^a$ are complex coordinates on the special
K\"ahler base $M_{\rm  SK}^{2\nh-2}$.\footnote{In \eqref{FS} we display the ungauged metric.
If some of its isometries are gauged the corresponding ordinary derivatives are replaced by appropriate covariant derivatives.}
The metric $g_{a\bb}(z,\bz)$ on
$M_{\rm  SK}^{2\nh-2}$ satisfies~\eqref{gdef} and thus can be
characterized by a holomorphic prepotential $\cG(z)$.\footnote{For type IIA, we denote the coordinates and prepotential of $M_{\rm SK}^{2\nh-2}$ in the hyper-sector by $z^a$ and $\cG$, respectively, in order to distinguish it from the coordinates $t^i$ and prepotential $\cF$ in the vector multiplet sector. For type IIB the r\^oles of $t^i$ and $z^a$ and $\cF$ and $\cG$ will be reversed.}  
The couplings $\cN_{IJ}(z,\bz)$ are given in terms of  $\cG$ as
\begin{equation}\label{Ndef}
   \cN_{IJ} = - \I \bar{\cG}_{IJ} + 2\, \frac{\im \cG_{IK} z^K\, \im
      \cG_{JL} z^L}{\im \cG_{MN} z^M z^N} \ , 
\end{equation}
with $z^I=(\I,z^a)$. In type II string theory reduced on, in
particular, a Calabi--Yau manifold, $\phi$ and $\si$ are identified
with the four-dimensional dilaton and universal axion, respectively,
while $\xi^I$ and $\txi_I$ are identified with four-dimensional scalar
fields arising from the RR sector. 

\subsection{$\SU(3)$ compactifications}\label{stt} %%%%%%%%%%%%%%%%%%%%

Let us now briefly discuss the notion of an $\SU(3)$
compactification. Recall first the moduli that appear when the
compactification space $X_6$ is a Calabi--Yau manifold. One finds that
the scalar fields in vector- and hypermultiplets arise from
$h^{(1,1)}$ deformations of the complexified K\"ahler form $J+\I B$ and from
$h^{(1,2)}$ deformations of the complex structure, or equivalently
deformations of the holomorphic 3-form $\Omega$. Since these
moduli can be varied independently, their moduli space is locally a
product 
\begin{equation}
  \label{CYmoduli}
  \M(X_6) \ =\ {\M}_{J} \times {\M}_{\Omega}\ ,
\end{equation}
with each component being a special-K\"ahler manifold.
Their respective K\"ahler potentials are given, prior to $\alpha'$ and
string loop corrections, by~\cite{CdO}  
\beq\label{KCY}
K_J=-\ln\int_{X_6} J\wedge J\wedge J\ ,\qquad
K_\Omega=-\ln \ii \int_{X_6} \Omega\wedge\bar\Omega\ .
\eeq
In type IIA compactifications one has ${\M}_{J}={\M}_{\rm  v}$ while
${\M}_{\Omega}=M_{\rm SK}^{2h^{(1,2)}}$ is the special-K\"ahler base
in the c-map \eqref{c-map}. The full space ${\M}_{\rm  h}$ also
includes the dilaton and axion plus $2h^{(1,2)}+2$ scalars
from the RR sector which parameterize the torus fibre.
In type IIB  compactifications the situation is reversed
and one has ${\M}_{\Omega}={\M}_{\rm  v}$ while
${\M}_{J}=M_{\rm SK}^{2h^{(1,1)}}$ is the special-K\"ahler base. 
The dilaton and axion plus $2h^{(1,1)}+2$ scalars
from the RR sector again complete ${\M}_{\rm  h}$.

In this paper we will be considering the weaker case where we only
assume that $X_6$ admits an $\SU(3)$ structure~\cite{waldram}. This means
we can still find a globally defined fundamental 2-form $J$ and an
almost complex structure with a corresponding globally defined complex
$(3,0)$-form $\Omega$. However, they are now no longer closed, so that 
\beq
\label{int-tor}
\dd J\neq 0 \ ,\qquad \dd\Omega\neq 0\ ,
\eeq  
and one says that the intrinsic torsion of the $\SU(3)$ structure on
$X_6$ is non-zero. This lack of integrability means that the space of
generic $\SU(3)$ structures on $X_6$ is infinite-dimensional, though
it nonetheless still decomposes as a product of special-K\"ahler
manifolds~\cite{GLW1,GLW2}, with K\"ahler potentials given
by~\eqref{KCY}. In the corresponding `low-energy' effective action the
intrinsic torsion plays the role of gauge charges and/or mass terms.
As a consequence, the four-dimensional supergravity is gauged.

For a realistic theory, we would like to identify a finite-dimensional subspace of $\SU(3)$ structures, as we had for the Calabi--Yau case. This should correspond to identifying some light ``moduli'' fields in the Kaluza--Klein spectrum of all possible deformations of $J$ and $\Omega$. In this paper we will simply assume such an expansion exists and then investigate the consequences for quantum corrections to the corresponding effective action. Our assumptions are (see~\cite{GLMW,GLW1,GLW2,KPM} for details): 
\begin{enumerate}
\item 
There is a finite-dimensional subspace $\M$ of $\SU(3)$ structures
such that we have the expansion
\begin{equation}
\label{expand}
   J + \ii B = t^i \omega_i \ , \qquad
   \Omega = z^I \alpha_I - \mathcal{G}_I(z) \beta^I\ .
\end{equation}
In contrast to the Calabi--Yau case, the lack of
integrability~\eqref{int-tor} means that the basis two-forms
$\omega_i$ and the dual four-forms $\tilde{\omega}^i$ as well as the
three-forms
$\alpha_I$ and $\beta^I$ are no longer  necessarily harmonic; these forms constitute a
closed set under the Hodge star and also under the exterior
derivative, namely  
\begin{equation}\label{diffcond2}
\begin{aligned}
\dd \alpha_{I} & \sim e_{Ii}\tilde\omega^i\ , & \qquad
\dd\beta^I &\sim p_{i}^I\tilde\omega^i , \\
\dd\omega_{i} &\sim p_i^I \alpha_I - e_{iI}\beta^{I} \ , & \qquad 
\dd\tilde\omega^{i} &\sim 0 \  ,
\end{aligned}
\end{equation}
where $e_{iI}$ and $p_{i}^I$ are constant matrices.

\item 
There are no additional light spin-3/2 fields other than the
$\mathcal{N}=2$ gravitini (this corresponds to an absence of moduli of 
type $(1,0)$ with respect to the almost complex
structure~\cite{GLW1,GLW2}).
\item 
There is an expansion around the large-volume limit. The moduli space manifestly includes the volume modulus $V$ since this is a rescaling of the real parts of $t^i$, here we in addition assume that the moduli are light compared with other Kaluza--Klein modes in the large-volume region.  
\end{enumerate}
This set-up can be viewed as the $\SU(3)$ structure analogue of a Calabi--Yau compactification with flux. The flux generates a potential for the Calabi--Yau moduli, but in the large volume limit this potential is generically small compared to the masses of the Kaluza--Klein modes. As a result the low-energy theory is just a gauged $\mathcal{N}=2$ supergravity theory for the moduli fields. Here we are requiring the existence of a family of $\SU(3)$ structures which are ``close'' to satisfying the Einstein equations, in the sense that the Ricci curvature is small compared to the Kaluza--Klein scale in the large volume limit. 

The existence of an $\SU(3)$ structure implies that the effective field theory in four dimensions is $\mathcal{N}=2$ supersymmetric. Given our assumptions,
the moduli space is again a product
\begin{equation}
  \M = {\M}_{J} \times {\M}_{\Omega}
\end{equation}
of a moduli space $\M_J$ of deformations of $J+\ii B$ and of a moduli
space $\M_\Omega$ of deformations of $\Omega$, where the volume modulus
$V$ is part of $\M_J$. Furthermore, the expansion~\eqref{expand} is such that the special K\"ahler structures on the infinite dimensional space of all $\SU(3)$-structures restrict to special K\"ahler structures on each component $\M_J$ and $\M_\Omega$ with K\"ahler potentials again given by~\eqref{KCY}. In addition, prior to including $\alpha'$ and string loop corrections, the low-energy effective theory is a gauged $\mathcal{N}=2$ supergravity theory. 

The RR potentials are assumed to admit an expansion in the same bases,
namely $\alpha_I$ and $\beta^I$ in type IIA and $\omega_i$,
$\tilde{\omega}^i$ and the volume form $\varepsilon$ in type IIB, 
so that
\begin{equation}
\label{RR-scalar}
\begin{gathered}
   C_3 = \xi^I \alpha_I + \tilde{\xi}_I \beta^I\ , \\
   C_0 = \xi^0\ , \qquad C_2 = \xi^i \omega_i\ , \qquad
   C_4 = \tilde{\xi}_i \tilde{\omega}^i\ , \qquad 
   C_6 = \tilde{\xi}_0 \varepsilon\ . 
\end{gathered}
\end{equation}
In Calabi--Yau compactifications the metric on the quaternionic
manifold $M_{\text{h}}$ 
takes the Ferrara--Sabharwal
form~\eqref{FS} to leading order. Related to the torus fibre the metric has
$2n_{\rm h}+1$ (perturbative) Peccei--Quinn shift symmetries
which read  
\beq\label{PQ}
\sigma \to \sigma + \gamma + c^I \tilde \xi_I\ ,\qquad \xi^I \to \xi^I + c^I\ ,\qquad \tilde\xi_I \to \tilde\xi_I + \tilde c_I\ ,\qquad \gamma, c^I, \tilde c_I \in \mathbb{R}\ .
\eeq
These symmetries arise from large gauge transformations of the type II $p$-form fields
$C_p$, that is, shifting them by constant multiples of the (closed)
basis forms $\omega_i$ etc. They are broken to
discrete shift symmetries by non-perturbative space-time physics
but nevertheless
imply a perturbative 
non-renormalization theorem in that perturbative corrections
of the hypermultiplets only occur at one-loop but not beyond 
\cite{Strominger,Antoniadis:1997eg,Gunther:1998sc,AMTV,loop}.

By contrast, for $\SU(3)$ structures the basis forms
are generically no longer harmonic, so in general the shift
symmetries \eqref{PQ}
are broken and the corresponding scalar fields are
massive. Nonetheless, simple dimensional analysis shows that the
leading-order calculation of the kinetic terms does not see the
derivatives of the basis forms and so the metric on the hypermultiplet
space still takes the Ferrara--Sabharwal form~\eqref{FS}. At higher
order in $\alpha'$ and $g_s$ one would naively expect that corrections
to the metric see the fact that the Peccei--Quinn symmetries are broken.

In summary, our assumptions imply that at leading order -- just as
in the Calabi--Yau case -- in type IIA one has ${\M}_{J}={\M}_{\rm v}$
while ${\M}_{\Omega}$ is the base of a special quaternionic-K\"ahler
manifold ${\M}_{\rm  h}$. In type IIB the situation is reversed
and one has  ${\M}_{\Omega}={\M}_{\rm  v}$ while ${\M}_{J}$ is the base of
${\M}_{\rm  h}$. In both cases the special quaternionic manifold
${\M}_{\rm  h}$ includes the dilaton and axion.

%%%%%%%%%%%%%%%%%%%%%%%%%%%%%%%%%%%%%%%%%%%%%%%%%%%%%%%%%%%%%

\section{Perturbative $\alpha'$ and $g_s$ corrections}

In this section we discuss the structure of the perturbative $\alpha'$
and $g_s$ corrections to the moduli spaces metrics in the class of
$SU(3)$ structure compactifications defined above, focussing on the
leading-order contributions. The key point is that they are strongly constrained by the $\mathcal{N}=2$ supersymmetry which by construction survives the compactification. In essence we find that several of the usual arguments that apply to Calabi--Yau compactifications go through in this case too.

First note that if the low energy spectrum contains no massive spin-3/2
multiplets, $\mathcal{N}=2$ supersymmetry enforces the split
\eqref{VHsplit} into locally independent special-K\"ahler and
quaternionic moduli spaces, which thus has to persist after including
perturbative and non-perturbative $\alpha'$ and $g_s$ corrections. The
$\alpha'$ corrections enter in an expansion in the inverse of the
volume modulus $V^{-1}$, while the loop corrections are in powers of
the four-dimensional dilaton
$\e^{2\phi}=V^{-1}\e^{2\varphi}$. However, in both type II
compactifications the dilaton is part of a hypermultiplet and thus the
component $M_{\rm h}$ receives quantum corrections, whereas $M_{\rm
  v}$ is uncorrected and thus ``exact''  already at the string tree
level.\footnote{Here \emph{exact} is in quotation marks since at
  special points in the moduli space -- for example at the conifold
  point -- non-perturbative corrections which are not governed by the
  dilaton can occur and correct $M_{\rm v}$ \cite{Strominger:1995cz}.}
The volume modulus on the other hand, resides in $M_J$. Therefore, in
type IIA $\alpha'$ corrections appear in $M_{\rm v}$ whereas in type
IIB they correct $M_{\rm h}$. This situation is identical to the
situation for Calabi--Yau compactifications and is summarized in
Table~\ref{corr}. 

It is important to note that for the leading order corrections there
are generically mixing terms of the form
$\partial_\mu\phi\,\partial^\mu\ln V$ between the dilaton and the
volume. For Calabi--Yau compactifications, the tree-level coefficent
of this term has been computed in \cite{Becker:2002nn,AMTV} but the
one-loop correction has only been inferred indirectly in \cite{AMTV}
by insisting that the split \eqref{VHsplit} continues to hold at the
loop level. To diagonalize such terms, one requires mixing between the
dilaton and volume modulus in defining the combinations that appear in
the appropriate vector and hypermultiplet moduli spaces. Such mixing
cannot affect the leading-order contributions to the moduli space
metrics but will enter at higher order.  

\begin{table}[ht]
   \begin{center}
      \bigskip
      \begin{tabular}{|c|c|c|c|c|} \hline \rule[-0.3cm]{0cm}{0.8cm}
     corrections & $M_{\rm v}^{\rm IIA}$ & $M_{\rm h}^{\rm IIA}$ &
     $M_{\rm v}^{\rm IIB}$ & $M_{\rm h}^{\rm IIB}$ \\ \hline 
         \rule[-0.3cm]{0cm}{0.8cm} $\alpha'$&
        yes &no&no&yes \\ \hline
         \rule[-0.3cm]{0cm}{0.8cm} {$g_s$} & 
         no &yes&no&yes \\ \hline
            \end{tabular}
      \caption{Structure of $\alpha'$ and $g_s$ corrections in type IIA 
         and type IIB.}
      \label{corr}
   \end{center}
\end{table}

In the following, we first consider only the metric moduli $\re t^i$
and $z^a$ and  use the splitting \eqref{VHsplit} and our assumptions
about the expansion~\eqref{expand} to constrain the form
of the leading corrections to the metrics on $M_{\rm v}$ and $M_{\rm h}$. 
We then consider the dependence of the NS $B$-field and RR moduli,
using analogues of the standard Calabi--Yau compactification arguments
to investigate how first the full perturbative vector-multiplet $\alpha'$
corrections and $g_s$ corrections are constrained.

%%%%%%%%%%%%%%%%%%%%%%%%%%%%%%%%%%%%%%%%%%%%%%%%%%%%%%%%%%%%
\subsection{Constraints on the metric moduli}\label{CKK}
%%%%%%%%%%%%%%%%%%%%%%%%%%%%%%%%%%%%%%%%%%%%%%%%%%%%%%%%%%%%

Let us first focus on the leading perturbative corrections both in
$\alpha'$ and $g_s$ for the metric moduli $\re t^i$ and $z^a$. One can
distinguish two types of terms. One is the contribution of pointlike
massive string states, either in tree-level exchange or in one-loop
diagrams. Given that we are expanding around a large volume limit of the compactification space, the masses of these states are much larger than the inverse radius of $X_6$. Hence they are intrinsically ten-dimensional effects and correspond to a dimensional reduction of the ten-dimensional string tree-level and one-loop effective action. The second type of term is intrinsically four-dimensional and give  ``threshold effects'' arising from integrating out heavy states coming from Kaluza--Klein modes and wound string states on $X_6$. For type II Calabi--Yau compactifications one notably finds that all such threshold corrections in fact vanish~\cite{Antoniadis:1997eg}. 

Before going into the details of the $\alpha'$ and $g_s$ corrections of the four-dimensional theory, to be discussed in the next subsections, let us recall the structure of the higher order string tree-level and one-loop terms
for the gravitational modes of the ten-dimensional effective action.
They have the form (see e.g.~\cite{Antoniadis:1997eg,GGV}) 
\begin{equation} \label{L_IIAB}
\begin{split}
  \cL_\text{IIA}|_{R^4} &
    \sim \zeta(3)\, \e^{-2\varphi} (t_8 t_8 + \tfrac{1}{4}\ep\ep)\+ R^4 
       + 2\zeta(2)\+ (t_8 t_8 - \tfrac{1}{4}\ep\ep)\+ R^4 + B \wedge I_8 \ , 
\\[.5ex]
  \cL_\text{IIB}|_{R^4} &
     \sim \zeta(3)\, \e^{-2\varphi} (t_8 t_8 + \tfrac{1}{4}\ep\ep)\+ R^4 
        + 2\zeta(2)\+ (t_8 t_8 + \tfrac{1}{4}\ep\ep)\+ R^4\ ,
\end{split}
\end{equation}
where $\varphi$ is the ten-dimensional dilaton, $\zeta$ is the Riemann zeta function, and
\begin{equation} 
\begin{aligned}\label{tdef}
   (t_8 t_8 \pm \tfrac{1}{4}\ep\ep)\+ R^4 
      \equiv \big( t^{M_1\cdots M_8}t_{N_1\cdots N_8} 
            &\pm \tfrac14 \ep^{PQM_1 \cdots M_8} 
               \ep_{PQN_1\cdots N_8} \big) \\ & \qquad 
            \times R_{M_1M_2}{}^{N_1N_2} 
            \cdots R_{M_7M_8}{}^{N_7N_8}\ .
\end{aligned}
\end{equation}
$R_{M_1M_2}{}^{N_1N_2}$ denotes the Riemann tensor, $\epsilon$ is the totally antisymmetric Levi--Civita tensor, and $t_8$ is antisymmetric in each successive pair of indices and symmetric under the exchange of any two pairs. Given an
antisymmetric tensor $M_{MN}$, 
the $t$-tensor contraction reads
\begin{equation}
 t_{M_1\cdots M_8} M^{M_1M_2}\cdots M^{M_7M_8} = 24 \tr M^4 - 6 (\tr M^2)^2\ .
\end{equation}
We also have
\begin{equation}\label{I8}
 4608\+ (2\pi)^4\, I_8(R) \equiv t_{N_1\cdots N_8} R^{N_1N_2} \wedge  \cdots \wedge  R^{N_7N_8} = \big[ 24\tr R^4 - 6\+ (\tr R^2)^2 \big]\ ,
\end{equation}
where $R^{N_1N_2}$ is the curvature 2-form. Note that the supersymmetric completion of these terms will include, in the NSNS sector, higher derivative objects built from curvatures, $H=\dd B$ and derivatives of the dilaton, which would be relevant if for instance one was interested in backgrounds with non-trivial $H$-flux. However, although recently some significant progress has been made~\cite{Liu:2013dna}, less is known about the exact form of these terms.   

\subsubsection{$\alpha'$ corrections}
\label{sec:alphapmetric}

Let us start by considering the leading $\alpha'$ corrections. 
% which arise 
As usual, all the contributions arise
from the ten-dimensional effective
action \eqref{L_IIAB}.\footnote{Note that we are expanding in terms of supersymmetry representations. Since we are off-shell these do not necessarily correspond to eigenstates of the Laplacian, and so, in contrast to the conventional case, there are potentially threshold corrections from tree-level exchange of Kaluza--Klein modes. However, by assumption the zeroth-order low-energy effective theory is a gauged supergravity for the moduli alone. This implies that the $\SU(3)$ background does not source the heavy Kaluza--Klein modes, and hence such contributions are absent. For the string winding modes, conservation of winding number means they similarly cannot be sourced. By contrast, both types of heavy modes can contribute in loops.} In principle, this can be derived by directly performing a Kaluza--Klein reduction of \eqref{L_IIAB}. 
%At the two-derivative level they lead to corrections of the Einstein
%term and the two scalar field metrics just as in the Calabi--Yau
%case~\cite{Antoniadis:1997eg,Becker:2002nn,AMTV}. Let us start by
%considering the Einstein term; 
Since $t_8t_8R^4$ never involves
contractions of indices on the same Riemann tensor, in the
four-dimensional effective action any correction to
the Einstein term must come from $\ep\ep
R^4$~\cite{Antoniadis:1997eg}. More precisely, the only contraction
that gives a term with only two space-time derivatives is where $PQ$
in~\eqref{tdef} are four-dimensional space-time indices, and the two
other space-time indices on each $\ep$ contract the  same Riemann
tensor. (If they contract different Riemann tensors one obtains
higher-derivative scalar field terms.) Integrating over the internal
space, we obtain 
\begin{equation}\label{Eulerreduction}
\begin{aligned}
     \int_{X_6} & \ep^{\rho\sigma\mu_1\mu_2}\ep^{m_1\cdots m_6}
        \ep_{\rho\sigma\nu_1\nu_2}\ep_{n_1\cdots n_6}
           R_{\mu_1\mu_2}{}^{\nu_1\nu_2}
           R_{m_1m_2}{}^{n_1n_2}R_{m_3n_4}{}^{n_3n_4}R_{m_5n_6}{}^{n_5n_6}
           \\[-.5ex]
        &\sim \mathcal{R}\int_{X_6}(\ep^{m_1\cdots m_6}\ep_{n_1\cdots n_6}
           R_{m_1m_2}{}^{n_1n_2}R_{m_3n_4}{}^{n_3n_4}R_{m_5n_6}{}^{n_5n_6})
        \sim \mathcal{R}\, \chi(X_6) \ , 
\end{aligned}
\end{equation}
where $\mathcal{R}$ is the four-dimensional Ricci scalar and
$\chi(X_6)$ the Euler characteristic of $X_6$. Thus we see that for
$\SU(3)$ structures, the correction to the Ricci scalar term comes
only from $\ep\ep R^4$ and is proportional to $\chi$ exactly as for
Calabi--Yau compactifications.\footnote{The fact that this property does not only hold for Calabi-Yau
compactifications  has also been observed in \cite{CKP}.}

For the scalar kinetic energy corrections, we note that $\ep\ep R^4$
necessarily has four space-time indices and thus cannot
contribute. One can, however, get a correction from the $t_8t_8 R^4$
terms~\cite{Antoniadis:1997eg,Becker:2002nn,gemmer}. We will denote these
corrections to the metrics on the $\M_J$ and $\M_\Omega$ moduli spaces
as $\delta g^{\text{tree}}_J$ and $\delta g^{\text{tree}}_\Omega$, respectively, and the leading
order, uncorrected metrics as $g_J^0$ and $g_\Omega^0$.  Inspecting the terms in \eqref{L_IIAB}, we can now write the leading $\alpha'$-correction to both the IIA and IIB Lagrangians in the string frame as~\cite{AMTV}
\begin{equation}
\label{para-tree}
\begin{aligned}
  \Delta\cL_{\text{tree}}
    & \sim \e^{-2\phi} \Big( 1
         +  V^{-1} c \Big) \mathcal{R} \\*[-.5ex]
         & \qquad \qquad
         + \e^{-2\phi} \Big(\re g_J^0 
             + V^{-1}\re \delta g^{\text{tree}}_J
                \Big)_{\!ij}\, 
            \partial_\mu v^i \partial^\mu v^j \\*[-.5ex]
         & \qquad \qquad
         +  \e^{-2\phi} \Big( g_\Omega^0 
             +  V^{-1} \delta g^{\text{tree}}_\Omega 
                \Big)_{\!a\bb}\, 
            \partial_\mu z^a  \partial^\mu \bar{z}^{\bb} 
         + \dots\ ,
\end{aligned}
\end{equation}
where $v^i=\re t^i$, $\phi$ is the four-dimensional dilaton defined as
$\e^{2\phi}=V^{-1}\e^{2\varphi}$ and  
%As we have seen, the Ricci scalar correction is identical to that in
%the Calabi--Yau case. Hence, we can use the standard Calabi--Yau
%result to write 
\beq
\label{c-def}
   c = \frac{2\zeta(3)}{(2\pi)^3}\, \chi(X_6)\ .
\eeq

The next step is to Weyl-rescale to the Einstein frame and then expand in terms of $\e^{2\phi}$ and the inverse volume $V^{-1}$, which parametrize loop and $\alpha'$ corrections, respectively.\footnote{Just to reiterate, the definitions of both the four-dimensional dilaton $\phi$ and the volume $V$ change due to their mixing just discussed \cite{Becker:2002nn,AMTV}. However this, and any more general mixing that might appear for generic $\SU(3)$ backgrounds, does not change the corrections to the moduli space metrics at this order.} To leading order in $V^{-1}$, one finds
\begin{equation}
\label{einstein-tree}
\begin{aligned}
  \Delta\cL_{\text{tree}}
    & \sim \mathcal{R} 
         + \Big( \re g_J^0 
             + V^{-1}\big( 
                \delta g^{\text{tree}}_J - c \re g_J^0 \big)
                \Big)_{\!ij}\, 
            \partial_\mu v^i \partial^\mu v^j \\*[-.5ex]
         & \qquad \qquad
         + \Big( g_\Omega^0 
             + V^{-1}\big( \delta g^{\text{tree}}_\Omega 
                - c\, g_\Omega^0\big)
                \Big)_{\!a\bb}\, 
            \partial_\mu z^a  \partial^\mu \bar{z}^{\bb} 
         + \dots\ .
\end{aligned}
\end{equation}
Recalling that there can be no $\alpha'$ corrections to the complex structure moduli space we immediately have 
\begin{equation}
\label{gOmega-tree}
   \delta g^{\text{tree}}_\Omega = c\, g_\Omega^0\, .
\end{equation}
One might be tempted to argue, for example by invoking a putative
mirror symmetry, that $\re \delta g_J$ is similarly proportional to $\re g_J^0$, as it is in the Calabi--Yau case. In the next section, by considering the contributions of the imaginary parts of the moduli coming from the NS $B$-field, we will see that this is indeed the case.

\subsubsection{$g_s$ corrections}\label{gscorr}

For loop corrections both ten-dimensional and threshold contributions
are relevant, and -- a priori -- there is little one can deduce about the
form of the four-dimensional low-energy action. Before Weyl-rescaling the loop-corrected Lagrangians have the form 
\begin{equation}
\label{para-loop}
\begin{aligned}
  \Delta\cL_{\text{loop}}^{\text{IIA/IIB}}
    & \sim \Big( V\e^{-2\varphi} 
         + f^{\text{A/B}} \Big) \mathcal{R} % \\*[-.5ex]
%          & \qquad \qquad
         + \Big( V\e^{-2\varphi} \re g_J^0 
             + \re \delta g^{\text{loop,A/B}}_J
                \Big)_{\!ij}\, 
            \partial_\mu v^i \partial^\mu v^j \\*[-.5ex]
         & \qquad \qquad
         + \Big( V\e^{-2\varphi} g_\Omega^0 
             + \delta g^{\text{loop,A/B}}_\Omega 
                \Big)_{\!a\bb}\, 
            \partial_\mu z^a  \partial^\mu \bar{z}^{\bb} 
         + \dots\ ,
\end{aligned}
\end{equation}
where 
$\delta g^{\text{loop,A/B}}_J$ and $\delta g^{\text{loop,A/B}}_\Omega
$ denote the leading order loop corrections of the two metrics while 
$f^{\text{A/B}}$ parametrize the loop-correction to the Einstein term.
All three corrections contain a contribution from the reduction of the
ten-dimensional terms given in \eqref{L_IIAB} and in principle an
additional threshold correction.
After rescaling to the Einstein frame one finds
\begin{equation}
\label{einstein-loop}
\begin{aligned}
  \Delta\cL_{\text{loop}}^{\text{IIA/IIB}}
    & \sim \mathcal{R} 
         + \Big( \re g_J^0 
             + \ee^{2\phi}\big( 
                \re\delta g^{\text{loop,A/B}}_J 
                   - f^{\text{A/B}} \re g_J^0 \big) \Big)_{\!ij}\, 
            \partial_\mu v^i \partial^\mu v^j \\*[-.5ex]
         & \qquad \qquad
         + \Big( g_\Omega^0 
             + \ee^{2\phi}\big(\delta g^{\text{loop,A/B}}_\Omega 
                - f^{\text{A/B}} g_\Omega^0\big)
                \Big)_{\!a\bb}\, 
            \partial_\mu z^a  \partial^\mu \bar{z}^{\bb} 
         + \dots\ .
\end{aligned}
\end{equation}
The requirement that there be no loop corrections to the vector multiplet moduli means that
\begin{equation}
\label{no-loop}
   \re \delta g^{\text{loop,A}}_J = f^{\text{A}} \re g_J^0 \ , 
   \qquad
   \delta g^{\text{loop,B}}_\Omega = f^{\text{B}} g_\Omega^0 \ . 
\end{equation}
%
%As stands we can put no further constraints on the corrections. 
Since $g_J$ can only depend on the $v^i$ and  $g_\Omega$ can only
depend on the $z^a$ we can infer  that $f^{\text{A}}$ is a function of only $v^i$ and $f^{\text{B}}$ is a function of only $z^a$.

At this point we can put no further constraints on the corrections. 
However, it is interesting to note that for Calabi--Yau compactifications the
type IIA and IIB corrections are related by 
\begin{equation}
\label{odd-even-rel}
   f^{\text{A}} = - f^{\text{B}} \ , \qquad
   \re \delta g^{\text{loop,A}}_J = \re \delta g^{\text{loop,B}}_J \ , \qquad
   \delta g^{\text{loop,A}}_\Omega = \delta g^{\text{loop,B}}_\Omega  \ . 
\end{equation}
These relations arise by
%come from 
noting from which worldsheet spin-structure
sectors the terms originate~\cite{Antoniadis:1997eg}. The Einstein term
corrections come only from the odd-odd sector, while the metric
corrections come from the even-even sector. 
%In the bulk theory, these sectors correspond to the signs under
%parity transformations acting separately in the two factors of the
%local $O(9,1)\times O(9,1)$ symmetry. 
Since $t_8$ is even under parity, one sees that in the ten-dimensional action~\eqref{L_IIAB}, the one-loop $\epsilon\epsilon$ term is odd-odd, and changes sign between IIA and IIB, while the $t_8t_8$ term is even-even and is the same for both IIA and IIB. 

Using \eqref{Eulerreduction}, we see that the loop corrections which arise from solely reducing the terms in the ten-dimensional action given in \eqref{L_IIAB} do respect the relations~\eqref{odd-even-rel} even for $\SU(3)$-structure backgrounds. Therefore, with the additional assumption that the same is true for the threshold contributions, one obtains
\begin{equation}
\label{loop-conj}
   \re \delta g^{\text{loop,A}}_J = \re \delta g^{\text{loop,B}}_J 
      = f \re g^0_J\ , \qquad
   \delta g^{\text{loop,A}}_\Omega = \delta g^{\text{loop,B}}_\Omega  
      = - f g^0_\Omega \ , 
\end{equation}
where $f=f^{\text{A}}=-f^{\text{B}}$. Furthermore, since the corrections to $\re g_J$ can only depend on the moduli $\re t^i$ and the corrections to $g_\Omega$ can only depend on the moduli $z^a$, we also have that 
\begin{equation}
   f = \text{constant} \ . 
\end{equation}
Note that we would have come to the same conclusions by assuming that
there is a mirror symmetry between IIA and IIB sending $f$ to $-f$. In
fact, as we will see in section~\ref{pgs}, we are also led to something very close to the relations~\eqref{loop-conj} once we consider the form of the RR kinetic terms in the hypermultiplet sector, without needing the assumption about threshold corrections or mirror symmetry made here. 

%%%%%%%%%%%%%%%%%%%%%%%%%%%%%%%%%%%%%%%%%%%%%%%%%%%%%%%

\subsection{Perturbative $\alpha'$ corrections}

As discussed in section \ref{sec:alphapmetric}, while there are necessarily no $\alpha'$ corrections to $g_\Omega$, we could say nothing concrete about the corrections to $g_J$. However, we did not consider
higher-derivative couplings of the NS $B$-field as they are not yet completely known~\cite{Liu:2013dna}. Let us now address that issue and see how it might allow us to also constrain the perturbative corrections to $g_J$. 

In the reduction, the light modes of $B$ combine with the deformation
of $J$, as in~\eqref{expand}, to form complex scalar coordinates $t^i$
on $M_J$. The $\mathcal{N}=2$ prepotential $\cF$ that determines the
special-K\"ahler metric depends holomorphically on $t^i$ and is given
in the large volume limit -- so as to match~\eqref{KCY} -- by  
\begin{equation}\label{F0}
    \cF_0(t) = \kappa_{ijk}\,t^i t^jt^k\ ,
\end{equation}
with $\kappa_{ijk}$ real constants and we take $X^0=\ii$ and
$X^i=t^i$. (In Calabi--Yau compactifications the $\kappa_{ijk}$ are the
classical intersection numbers; more  generally they are related to
the basis forms $\omega_i$~\cite{GLW1,GLW2}.) 
In order to determine or constrain perturbative $\alpha'$ corrections,
we have to determine the
sub-leading corrections to $\cF$. 
Expanding in large $t^i$ we have generically
\begin{equation}\label{Fcomplete}
\cF(t) = \cF_0(t) + \alpha_{ij}t^it^j + \beta_i t^i + \gamma + \CF(t)\ ,
\end{equation}
where $\CF$ contains non-perturbative corrections 
(i.e.\ instanton corrections) together with possibly negative powers 
of $t^i$.
Here we use the fact that each power in the $\alpha'$ expansion
comes with a volume factor $V^{-1/3}$. From~\eqref{KCY} we see that
$V$ is cubic in $t^i$
\begin{equation}
   V = \tfrac{1}{6}\kappa_{ijk}(t+\bar t)^i (t+\bar t)^j(t+\bar t)^k 
      = \tfrac{1}{6}\e^{-K_0} \ ,
\end{equation}
where $K_0$ is the leading order K\"ahler potential computed from
$\cF_0$, and hence the corrections to $\cF_0$ are in descending
powers of $t^i$. 

Inserting \eqref{Fcomplete} into \eqref{gdef} one obtains  
\begin{equation}
\label{cubic-K}
   \e^{-K} =
      \kappa_{ijk}(t+\bar t)^i (t+\bar t)^j(t+\bar t)^k
      +a_{ij}t^i\bar t^j 
      +\ii b_i (t-\bar t)^i 
      + c + \ldots\ ,
\end{equation}
where 
\begin{equation}
   a_{ij}= 2 \ii (\alpha_{ij}-\bar\alpha_{ij})\ ,\qquad
   \ii b_i = (\beta_i-\bar\beta_i)\ ,\qquad
   c= 2\ii(\gamma-\bar\gamma)\ .
\end{equation}
So we see that the real parts of $\alpha_{ij}$, $\beta_i$
and $\gamma$ actually do not enter the K\"ahler potential and for our purpose 
may be
set to zero without loss of generality.\footnote{In Calabi--Yau compactifications they are determined to be non-vanishing using mirror symmetry, but this plays no role in the following.}
Computing the metric from \eqref{cubic-K} we find
\begin{equation}
\label{gJ-corr0}
\begin{aligned}
   g_{J\,i\bar{j}} &= -\,\e^{K}(6\kappa_{ij}+a_{ij})
     +  \e^{2K}(3\kappa_i +a_{ik}\bar t^k +\I b_i)(3\kappa_j +a_{jk}
        t^k -\I  b_j) \ , 
\end{aligned}
\end{equation}
where we abbreviated
\begin{equation}
\label{gJ-corra}
\begin{aligned}
\kappa_{ij} = \kappa_{ijk}(t+\bar t)^k \ ,\qquad
\kappa_i = \kappa_{ilk}(t+\bar t)^l(t+\bar t)^k\ .
\end{aligned}
\end{equation}

We would like to determine the values for $a_{ij}$, $b_{i}$ and $c$. Recall
that for Calabi--Yau compactifications there is a perturbative
Peccei--Quinn symmetry $B\to B+\lambda^i\omega_i$, which implies that
$g_{i\bar{j}}^J$ has an isometry $t^i\to t^i+\ii \lambda^i$ for 
constant real $\lambda^i$. This is very constraining, implying in
particular that $a_{ij}=b_i=0$. In fact, the PQ symmetry together with
the large volume limit is enough to also prove that no negative powers 
of $t^i$ can appear in $\cF$ and that the perturbative prepotential can only be
$\cF(t)=\cF_0(t)+\gamma$ with $\gamma$ constant.\footnote{To 
see this one inserts an arbitrary function into \eqref{gdef}, computes
the metric and imposes the PQ symmetry -- this determines
$\cF(t)=\cF_0(t)+\gamma$ (modulo terms that do not contribute to $K$). 
In particular, the term $\CF$ in \eqref{Fcomplete} cannot have 
negative powers of $t^i$ and only 
contains  non-perturbative corrections which
break the PQ symmetry to a discrete subgroup allowing for
the dependence $\CF(e^{2\pi t})$.}

Now let us turn to the general $\SU(3)$ structure case. In general the
PQ symmetry is a priori not present since the basis forms are not 
necessarily closed. However, as discussed in the previous sub-section,
the leading string corrections to supergravity, including those
involving $B$, appear at order $\alpha^{\prime3}$. Thus all leading
corrections to $g^J_{i\bar{j}}$ are suppressed by a factor of $V^{-1}$
and hence by cubic powers in $t^i$. This means, simply by counting
powers of $t$, that we expect $a_{ij}=b_{i}=0$ and in fact, somewhat
surprisingly, the leading order corrected metric still has the PQ symmetry.

In the analysis so far we made an implicit assumption
since the  expansion \eqref{Fcomplete} in $V^{-1/3}$ strictly
defines $\cF$ as a sum of functions homogeneous under the rescaling
$t^i\to\mu t^i$, that is
\begin{equation}
\label{homogen}
   \cF(t) = \cF_0(t) + \cF_1(t) + \cF_2(t) + \dots 
      + \cF_{np}(t)\ ,
\end{equation}
where $\cF_i(t)$ scales as $\mu^{3-i}$ and $\cF_{np}(t)$ is the
non-perturbative correction. Thus in~\eqref{Fcomplete} we are really
extracting only the polynomial parts of each $\cF_i(t)$ in the
expansion, relegating the non-polynomial contributions 
to $\CF$. The latter  would signal power-like singularities in the variables $t^i$ and naively one might expect their absence also in the $\SU(3)$ structure case, although notably, under T-duality, the contributions of wound strings are precisely of this type. For the leading $\alpha'$ corrections we can actually address this question directly. First, as for the $\alpha$ and $\beta$ terms above, since the leading string corrections to supergravity, including those involving the $B$-field, appear at order $\alpha^{\prime3}$, we can immediately argue that $\cF_1$ and $\cF_2$ vanish, since they correspond to $\alpha'$ and $\alpha^{\prime2}$ corrections. Turning to the $\alpha^{\prime3}$ correction,
$\cF_3$ we have already noted that the corrections to the kinetic terms are unaffected by moduli mixing to this order. Thus we can take the zeroth-order definitions, and identify $\im t^i$ with the expansion of the $B$-field in~\eqref{expand}. Next we note, just by power counting, that the leading higher-derivative corrections to the ten-dimensional effective action can include only finite powers of $H=\dd B$ up to $H^4$. Hence we can only have \emph{polynomial} dependence on $\im t^i$ in the correction to the metric $\delta g_J$. Together with the homogeneity condition this is enough to argue that in fact $\cF_3$ is constant.

To summarize, for any $\SU(3)$ structure compactifications we have
argued that the $\alpha'$-corrected prepotential on $M_J$
has the form 
\begin{equation}\label{Ffinal}
   \cF= \kappa_{ijk}\,t^i t^jt^k - \tfrac{1}{2}\ii\tilde{c} 
\end{equation}
where $\tilde{c}$ is constant. Note that this is exactly the same form as for Calabi--Yau compactifications, in which case $\tilde{c}=2\zeta(3)\chi(X_6)/(2\pi)^3$ is proportional to the Euler characteristic (and equal to $c$ in~\eqref{c-def}). The corresponding metric is given by 
\begin{equation}
\label{gJ-corr2}
\begin{aligned}
   g_{J\,i\bar{j}} &= -6\,\e^{K}\kappa_{ij}
+ 9\, \e^{2K}\kappa_i \kappa_j\ ,\quad {\rm with}\quad
   \e^{-K} =
      \kappa_{ijk}(t+\bar t)^i (t+\bar t)^j(t+\bar t)^k  + \tilde{c} \ ,
\end{aligned}
\end{equation}
so that expanding to first order in $V^{-1}$ gives
\begin{equation}
\label{gJ-corr3}
\begin{aligned}
   g_{J\,i\bar{j}} = g^0_{J\,i\bar{j}} + \e^{K_0}\delta g_{J\,i\bar{j}}
\end{aligned}
\end{equation}
where $g^{0}_{J\,i\bar{j}}$ and $K_0$ are the leading order metric and
K\"ahler potential computed from the cubic prepotential ${\cal F}_0$
given in \eqref{F0} and\footnote{\label{by-parts}Note that the second term appears to violate the conjecture in section~\ref{CKK} that the contributions to $\delta g^{\text{tree}}_J$ and $\delta g^{\text{tree}}_\Omega$ from the $t_8t_8 R^4$ term are proportional to $g_J^0$ and $g_\Omega^0$ respectively. However just as in the Calabi--Yau case~\cite{Becker:2002nn,AMTV}, the conjecture is really that these are the forms of the contributions up to pieces arising from a total derivative term in the expansion of $t_8t_8R^4$. The point is that integrating such a term by parts against the $V^{-1}\ee^{-2\phi}$ coefficient in~\eqref{para-tree} generates the second term in~\eqref{gJ-corr} for $\delta g_J$ but no such term for $\delta g_\Omega$. The corresponding term is also absent for $\delta g^{\text{loop}}_J$ in the reduction of the ten-dimensional loop correction, since there the coefficient is independent of $V$.}  
\begin{equation}
\label{gJ-corr}
 \delta g_{J\,i\bar{j}} =-\tilde{c}\, g^{0}_{J\,i\bar{j}} 
    - 9\tilde{c} \, \e^{2K_0}\kappa_i \kappa_j\ . 
\end{equation}
This result might seem somewhat surprising as we just argued that the PQ isometries are not present in $\SU(3)$ structure compactifications, yet the metric given in \eqref{gJ-corr2} does have them. This is arising because the ten-dimensional corrections have a very particular universal form, which together with $\cN=2$ supersymmetry imply the absence of 
subleading polynomial corrections in the prepotential. In that sense the PQ isometries can be viewed as ``accidental'' symmetries. (Of course they are broken in the potential already at leading order \cite{GLW1}.) 

%Before we turn to the hypermultiplet sector, 
%it is natural 
%one can ask what is 

The arguments used so far did not determine 
the value of $\tilde{c}$ for $\SU(3)$ structure compactifications. It
is 
%natural 
tempting to conjecture that it is again equal to $c$ in~\eqref{c-def}
and hence given by the Euler characteristic. This would be consistent
with any putative mirror symmetry given the
relation~\eqref{gOmega-tree}. It is also interesting to note that it
is consistent with the arguments of  \cite{KashaniPoor:2013en}, where it was shown that manifolds $X_6$ with vanishing Euler
characteristic necessarily have an additional $SU(2)$ structure
and the compactification can be viewed as a spontaneously broken $N=4$
theory. The $N=4$ supersymmetry then forbids any perturbative
corrections
and implies $\tilde{c}\sim \chi(X_6)$
or in other words  that $\tilde{c}$ must vanish as the Euler
characteristic goes to zero.\footnote{We thank H.~Triendl for pointing
  this out.} 
%By this analysis alone we can at least argue that $\tilde{c}$ must vanish as the Euler characteristic goes to zero. 

One might also wonder if, as in the Calabi--Yau case, this leading contribution actually gives the perturbative $\alpha'$ correction to all orders. If the argument that $\cF_3$ is polynomial in $\im t^i$ can be extended to all $\cF_n$ then homogeneity implies that all the higher $\cF_n$ with $n>3$ do indeed vanish. However, although one can again argue that the corresponding  higher-derivative corrections must be polynomial in $H$, the fact that there may be moduli mixing means that we cannot conclude that $\cF_n$ is generally polynomial in $\im t^i$. Thus, as stands, we cannot argue against higher-order corrections.

\subsection{Perturbative $g_s$ corrections}\label{pgs}

As we discussed above, for Calabi--Yau compactifications the 
zero-modes $\xi, \tilde\xi$ of the RR potentials $C_p$ together with
the axion $\sigma$ in the hypermultiplet moduli space admit 
$2n_{\rm h}+1$ (perturbative) Peccei--Quinn shift symmetries 
given in \eqref{PQ}.
In $SU(3)$ compactifications the situation is more involved
since generically some of the scalar fields become
massive, or in other words the number of the zero modes is reduced,
because the basis forms are no longer harmonic.
As discussed in~\cite{GLMW,GLW1,GLW2,KPM}, a subset of the shift
symmetries \eqref{PQ} turns local in that $c^I, \tilde c_I$ become space-time dependent and appropriate couplings to the gauge fields are induced. $\cN=2$ supersymmetry in turn demands a non-trivial scalar potential which lifts some of the flat directions corresponding to the ``non-zero'' but light modes mentioned above.
An argument along the lines of ref.~\cite{KashaniPoor:2005si} further
shows that these gauged isometries survive after including
perturbative and non-perturbative corrections. 

The number of gauged isometries depends on the specific structure 
of the non-trivial torsion, that is, the constants $e_{iI}$ and
$p_i^I$ in~\eqref{diffcond2}. Nevertheless, one can determine that 
for any $SU(3)$ compactification at least  $n_{\rm h}$ of the
isometries in \eqref{PQ} survive perturbatively, and only the question
which are gauged depends on the details of $X_6$~\cite{GLW2}. To
review this argument, note that in this case, for the RR scalars
in~\eqref{RR-scalar}, 
\begin{equation}
\begin{aligned}
   \dd(\xi^I \alpha_I + \tilde{\xi}_I \beta^I) 
      &= (\dd \xi^I) \alpha_I + (\dd\tilde{\xi}_I) \beta^I
         + (\xi^I e_{iI} + \tilde{\xi}_I p_i^I ) \tilde{\omega}^i \ , \\
   \dd(\xi^0+\xi^i \omega_i+\tilde{\xi}_i
       \tilde{\omega}^i+\tilde{\xi}_0 \varepsilon) 
      &= \dd\xi^0 + (\dd\xi^i) \omega_i 
         + (\dd\tilde{\xi}_i)\tilde{\omega}^i
         + (\dd\tilde{\xi}_0) \varepsilon 
         + (\xi^i e_{iI}) \alpha^I - (\xi^i p_i^I) \beta_I  \ . 
\end{aligned}
\end{equation}
This means that the RR field strengths depend explicitly on the
combinations $\xi^I e_{iI} + \tilde{\xi}_I p_i^I$ in type IIA and
$\xi^i e_{iI}$ and $\xi^i p_i^I$ in type IIB. Thus for IIB it is clear
that at most one loses the $n_h-1$ isometries $\xi^i\to\xi^i+c^i$. For
type IIA, one notes that $\dd^2=0$ implies
$e_{iI}p_j^I-p_i^Ie_{jI}=0$ and hence the vectors $Z_i=(e_{iI},p_i^I)$
span an isotropic subspace of the $2n_h$-dimensional symplectic space
spanned by $\alpha_I$ and $\beta^I$. Thus there can be at most $n_h$
linearly independent $Z_i$ and hence at most $n_h$ combinations 
$\xi^Ie_{iI}+\tilde{\xi}_Ip_i^I$ that appear explicitly in the RR
field strengths and hence have broken PQ symmetry. This result implies
that all hypermultiplets can be dualized to tensor multiplets, 
where the scalars which transform  as in \eqref{PQ} are replaced by
dual antisymmetric tensors~\cite{HKC}.\footnote{In type IIB the tensor multiplets already arise in the field basis which naturally occurs in the Kaluza--Klein reduction~\cite{Bohm:1999uk}.}

This property was used in~\cite{loop} to parameterize the possible
perturbative corrections in Calabi--Yau compactifications in terms of
one scalar function $\Delta(z)$. $\cN=2$ supersymmetry alone already
constrains the string loop-corrected scalar field spaces to be
quaternionic-K\"ahler. Our assumption of $\SU(3)$ structure further
implies that $M_{\rm h}$ is a torus fibration over a special-K\"ahler
manifold. However, because not all the RR isometries survive, the
metric on $M_{\rm h}$ no longer has to be `special
quaternionic-K\"ahler', or in other words the loop-corrected metric is
generically not of the Ferrara--Sabharwal form given in 
\eqref{FS} with merely a loop-corrected prepotential $\cG$.
However, the existence of at least $n_{\rm h}+1$ unbroken
translational isometries additionally constrains the form of the metric
\cite{Strominger,Gunther:1998sc,loop},
and this is best described in terms of the dual tensor multiplets.
For Calabi--Yau compactifications -- and as we just argued, also for
$SU(3)$ structure compactifications -- all hypermultiplets can be 
dualized to tensor multiplets, and as a consequence the constraints
determined in \cite{HKC,TV} apply. Using the additional property that
the dilaton organizes the $g_s$ expansion, we can repeat the analysis performed in \cite{loop} and arrive at the same result that the corrections to $M_{\rm h}$ are  determined by a single function $\Delta$, that is the imaginary part of a holomorphic function of the base-coordinates $z$ and is homogeneous of degree zero. Explicitly, the correction to \eqref{FS} has to be of the form \cite{loop}
\beq 
\begin{aligned} 
\label{RSVmetric}
  \cL & = \frac{1 + 2\Delta\e^{2\phi}}{1+\Delta\e^{2\phi}}\,
        (\partial_\mu\phi)^2 
     + \frac{1+\Delta\e^{2\phi}}{1+2\Delta\e^{2\phi}}\, \e^{4\phi}
        \big( D_\mu\si - \xi^I D_\mu\txi_I - \Delta \cA_\mu \big)^2  
     \\ & \tab 
     + (1 + \Delta\+ \e^{2\phi})\, g_{a\bb}\,
        \partial_\mu z^a\partial^\mu\bz^{\bb} 
     - \frac{\e^{4\phi}}{1 + \Delta\+ \e^{2\phi}}\, \tilde{\cA}_\mu \tilde{\cA}^\mu
     + \Delta\+ Y_{IJ}\, D_\mu\xi^I D^\mu\xi^J 
     \\ & \tab 
     - \e^{2\phi}\+ \cT^{IJ} \big( 
        D_\mu\txi_I - 2\I\+ \cM_{IK}\+ D_\mu\xi^K \big)
        \big( D^\mu\txi_J + 2\I\+ \bar{\cM}_{JL}\+ D^\mu\xi^L \big)\ ,
\end{aligned}
\eeq
where we included appropriate covariant derivatives.
$\cA_\mu=-\im\+ (\p_a \cK\+ \partial_\mu z^a)$ is the K\"ahler connection on the
special-K\"ahler base, $\tilde{\cA}_\mu=\re\+ (\p_a \Delta\+
\partial_\mu z^a)$, and $\cM_{IJ} (z,\bz,\phi), \cT_{IJ} (z,\bz,\phi)$ are
quantum deformations of $\cN_{IJ}$ (defined in \eqref{Ndef})
and $2\re\cN_{IJ}$, respectively. All these corrections are
proportional to powers of $\Delta$ and the precise expressions of
$\cM_{IJ} (z,\bz,\phi)$, $\cT_{IJ} (z,\bz,\phi)$ and the matrix
$Y_{IJ}(z,\bz,\phi)$ can be found in \cite{loop}. Here we have written
$z^a$ for the coordinates on the base, corresponding to type IIA. In
type IIB these would be replaced by $t^i$.

Again, we can compare this form of the metric directly with the loop
corrections to $M_J$ and $M_\Omega$ we discussed in
section~\ref{gscorr}. It provides strong constraints on their
form. Comparing \eqref{einstein-loop}
with the leading-order kinetic terms for $z^a$ in IIA and for $t^i$ in IIB in~\eqref{RSVmetric}, we see that 
\begin{equation}\label{g-loop}
   \delta g_\Omega^{\text{loop,A}} 
       = \left(\Delta^{\text{A}} + f^{\text{A}} \right) g_\Omega^0 \ , 
   \qquad   
   \delta g_J^{\text{loop,B}} 
       = \left(\Delta^{\text{B}} + f^{\text{B}} \right) g_J^0 \ ,  
\end{equation}
where $\Delta^{\text{A}}$ and $\Delta^{\text{B}}$ are the relevant
functions of $z^a$ and $t^i$ respectively. Again, we see that the
corrections are in fact proportional to the zeroth-order metrics,
consistent with the conjecture~\eqref{loop-conj}. Furthermore, given
the relations~\eqref{no-loop} and the fact that the corrections to $g_\Omega$ and $g_J$ can only depend on $z^a$ and $t^i$ respectively, we can also conclude that 
\begin{equation}
   f^{\text{A}} = \text{constant} \ , 
   \qquad
   f^{\text{B}} = \text{constant} \ .
\end{equation}
This implies that in both type IIA and IIB the renormalization of the Einstein term (including the threshold corrections) is simply given by a constant.

Also in the hypermultiplet sector we see accidental PQ~symmetries.
%If $\Delta$ is constant we see again that, 
Although the expansion is not generically compatible with preserving all $2n_{\rm h}+1$ PQ~symmetries, this is not realized in the correction to the
hypermultiplet moduli space -- the corrected metric \eqref{RSVmetric}
still preserves all the shift symmetries. Instead, the breaking is only realized in the mass terms. Furthermore, it was argued in~\cite{loop} that the tensor
multiplet structure and the dilaton expansion was enough to exclude
any further corrections beyond one-loop. Using the same logic here,
the implication is that for $\SU(3)$ structure compactifications there
is a non-renormalization theorem stating that the hypermultiplet
metric can only be corrected perturbatively at one-loop but not beyond.

As stands we are still left with arbitrary functions
$\Delta^{\text{A}}(z)$ and $\Delta^{\text{B}}(t)$. However, as discussed in section~\ref{CKK},
%the fact that supersymmetry restricted the form so greatly, that
%again $f^{\text{A}}$ and $f^{\text{B}}$ are constant, and 
a putative mirror symmetry points to the conjecture that the correction terms respect the same spin-structure symmetries as in Calabi--Yau compactifications and both $\Delta^{\text{A}}$ and $\Delta^{\text{B}}$ are in fact constant. If in addition we invoke the arguments of~\cite{KashaniPoor:2013en} we can further constrain the constant to be proportional to the Euler characteristic $\chi(X_6)$. We therefore close with the conjecture
\begin{equation}
  \Delta^{\text{A}} = - \Delta^{\text{B}} = \text{constant} \sim
  \chi(X_6)\ . 
\end{equation}
%
%Again we can invoke the arguments of~\cite{KashaniPoor:2013en} to constrain the constant to vanish with $X_6$ has vanishing Euler characteristic $\chi(X_6)$. The strongest conjecture would be that it is equal to $\frac{\zeta(2)}{\pi^3} \chi(X_6)$ just as in the Calabi--Yau case. 

\section{Summary and Outlook} %%%%%%%%%%%%%%%%%%%%%%%%%%%%%%%%%%%%%%%%%%

We have constrained the leading perturbative $\alpha'$ and $g_s$ corrections to
the moduli space metric arising in compactifications of type II
theories on six-dimensional $SU(3)$ structure manifolds $X_6$, subject to
some simple assumptions about the low-energy modes. 
We have shown that both the $\alpha'$ and $g_s$ leading corrections to the four-dimensional curvature
scalar are constant, with the tree level contribution 
given by the Euler characteristic $\chi(X_6)$.
Furthermore, 
 the leading tree-level correction of the K\"ahler moduli space metric
corresponds to a constant term $\tilde c$ in the prepotential $\cF$ (cf.~\eqref{Ffinal}).
For the loop corrections we argued that a non-renormalization theorem holds
in that the metric in the hypermultiplet sector is corrected at
one-loop
but receives no further perturbative correction. For the one-loop
corrections we could not show
that they coincide for type IIA and type IIB but they might differ
by a moduli dependent function $\Delta^{\text{A/B}}$
(cf.~\eqref{no-loop}, \eqref{g-loop}).
We summarize these results in Table \ref{summary}.\footnote{Note that here we are presenting these corrections after Weyl-rescaling to the Einstein frame, in constrast to the expressions in section \ref{CKK}. Also, the full tree-level correction to $\delta g_J$ is given by~\eqref{gJ-corr}. For simplicity, we use a shorthand of only writing the first term in~\eqref{gJ-corr} (see comment in footnote~\ref{by-parts}).}

\begin{table}[ht]
   \begin{center}
      \bigskip
      \begin{tabular}{|c||c|c||c|c|} \hline \rule[-0.3cm]{0cm}{0.8cm}
      & \multicolumn{2}{c||}{$\delta g_J$} & \multicolumn{2}{c|}{$\delta g_\Omega$} \\ \cline{2-5}
      & tree & loop & tree  & loop 
    \\ \hline 
         \rule[-0.3cm]{0cm}{0.8cm} IIA&
       $-\tilde c g_J^0$ &0 &0 &$\Delta^{\text{A}} g_\Omega^0$ \\ \hline
         \rule[-0.3cm]{0cm}{0.8cm} IIB & 
       $ -\tilde c g_J^0$ &$ \Delta^{\text{B}} g_J^0$&0 & 0 \\ \hline
            \end{tabular}
      \caption{Summary of metric corrections.}
      \label{summary}
   \end{center}
\end{table}

While our analysis leaves $\tilde c$ and $\Delta^{\text{A/B}}$ undetermined, we were able to gather a variety of arguments which point to the fact that they are exactly as in Calabi-Yau compactifications, namely that $\Delta^{\text{A/B}}$ is constant and all constants are proportional to the Euler characteristic
\begin{equation} \label{alleuler}
\tilde c\sim\Delta^{\text{A/B}}\sim \chi(X_6)\ .
\end{equation}
This would also imply that, even though the mass terms originating from the gaugings break some of the PQ shift symmetries, the perturbatively corrected kinetic terms actually possess \emph{all} shift symmetries. In fact, the derivation of \eqref{RSVmetric} in \cite{loop} shows that this property holds even for non-constant functions $\Delta^{\text{A/B}}$ under some assumptions on their pole structure.

A key open question is
how dependent these results are on our assumptions about the
low-energy modes, including the absence of light spin-3/2
particles. We hope nonetheless that this work provides a useful basis
point for studying the form of general corrections to
non-Calabi--Yau compactifications. 
We also have only considered the case of $SU(3)$ structure manifolds,
without background fluxes. An obvious extension of our results would
be to consider more general $SU(3) \times SU(3)$-structure manifolds,
and/or NSNS and RR fluxes. One could also try to investigate the form
of the non-perturbative corrections. We hope to make progress on these
issues in the near future. 

\section*{Acknowledgments} %%%%%%%%%%%%%%%%%%%%%%%%%%%%%%%%%%%%%%%%%%%%

We have benefited from conversations and correspondence with  
R.~Minasian, B.~Pioline, F.~Saueressig, H.~Triendl,
S.~Vandoren, P.~Vanhove and O~Varela. 
We also thank the referee of our paper for several helpful comments and for raising 
some interesting further points.
J.L.\ thanks the Weizmann Institute and the Theory Group at CERN 
for their kind hospitality during the final stages of this work.
This work was supported in part by the joint Network Grant DFG/LU/419/9-1 and  EPSRC EP/I02784X/1, the ERC Starting Grant 259133 -- ObservableString,
the German-Israeli Foundation for Scientific Research and Development
(GIF I-1-03847.7/2009), the I-CORE program of the Planning and Budgeting Committee and the Israel Science
Foundation (grant number 1937/12), the EPSRC Programme Grant ``New
Geometric Structures from String Theory'' EP/K034456/1 and the STFC
Consolidated Grant ST/J0003533/1.

\raggedright %\small


\begin{thebibliography}{99} %%%%%%%%%%%%%%%%%%%%%%%%%%%%%%%%%%%%%%%%%%%%


\bibitem{waldram}
   J.~P.~Gauntlett, N.~W.~Kim, D.~Martelli and D.~Waldram,
   ``Fivebranes wrapped on SLAG three-cycles and related geometry,''
   JHEP {\bf 0111} (2001) 018
   [\arXiv{hep-th/0110034}],

   J.~P.~Gauntlett, D.~Martelli, S.~Pakis and D.~Waldram,
   ``G-structures and wrapped NS5-branes,''
   Commun.\ Math.\ Phys.\  {\bf 247} (2004) 421
   [\arXiv{hep-th/0205050}], 

   J.~P.~Gauntlett, D.~Martelli and D.~Waldram,
   ``Superstrings with intrinsic torsion,''
   Phys.\ Rev.\ D {\bf 69}, 086002 (2004)
   [\arXiv{hep-th/0302158}].

\bibitem{GLMW}
   S.~Gurrieri, J.~Louis, A.~Micu and D.~Waldram,
   ``Mirror symmetry in generalized Calabi--Yau compactifications,''
   Nucl.\ Phys.\ B {\bf 654} (2003) 61
   [\arXiv{hep-th/0211102}],

   S.~Gurrieri and A.~Micu,
   ``Type IIB theory on half-flat manifolds,''
   Class.\ Quant.\ Grav.\  {\bf 20} (2003) 2181
   [\arXiv{hep-th/0212278}].

\bibitem{Hitchin}
   N.~Hitchin, 
   ``The geometry of three-forms in six and seven
   dimensions,'' 
   J.\ Diff.\ Geom.\ {\bf 55} (2000), no.3 547 
   [\arXiv{math.DG/0010054}], 

   N.~Hitchin, 
   ``Stable forms and special metrics,''
   in ``Global Differential Geometry: The Mathematical Legacy of Alfred
   Gray'', M.Fernandez and J.A.Wolf (eds.), 
   Contemporary Mathematics {\bf 288}, American Mathematical Society,
   Providence (2001) [\arXiv{math.DG/0107101}],

   N.~Hitchin, 
   ``Generalized Calabi--Yau manifolds,''
   Quart.\ J.\ Math.\ Oxford Ser.\  {\bf 54} (2003) 281
   [\arXiv{math.DG/0209099}],

   M.~Gualtieri, ``Generalized Complex Geometry,'' 
   Oxford University DPhil thesis (2004) [\arXiv{math.DG/0401221}].

\bibitem{Chris}
   C.M.~Hull,
   ``Generalised geometry for M-theory,''
   JHEP {\bf 0707}, 079 (2007)
   [\arXiv{hep-th/0701203}].

\bibitem{EGG}
   P.P.~Pacheco and D.~Waldram,
   ``M-theory, exceptional generalised geometry and superpotentials,''
   JHEP {\bf 0809}, 123 (2008)
   [\arXiv{0804.1362} [hep-th]].

\bibitem{GMPT2}
   M.~Gra\~na, R.~Minasian, M.~Petrini and A.~Tomasiello,
   ``Generalized structures of $N = 1$ vacua,''
   JHEP {\bf 0511} (2005) 020
   [\arXiv{hep-th/0505212}].

\bibitem{JW}
   C.~Jeschek and F.~Witt,
   ``Generalised $G_2$-structures and type IIB superstrings,''
   JHEP {\bf 0503} (2005) 053
   [\arXiv{hep-th/0412280}].

\bibitem{GLW1}
  M.~Gra\~na, J.~Louis and D.~Waldram,
   ``Hitchin functionals in $N = 2$ supergravity,''
   JHEP {\bf 0601} (2006) 008
   [\arXiv{hep-th/0505264}].

\bibitem{GLW2}  M.~Grana, J.~Louis and D.~Waldram,
   ``$\SU(3)\times\SU(3)$ compactification and mirror duals of
   magnetic fluxes,'' 
   JHEP {\bf 0704}, 101 (2007)
   [\arXiv{hep-th/0612237}].

\bibitem{GLSW}
  M.~Grana, J.~Louis, A.~Sim and D.~Waldram,
  ``$E_{7(7)}$ formulation of $N=2$ backgrounds,''
  JHEP {\bf 0907} (2009) 104
  [\arXiv{0904.2333} [hep-th]].
 
\bibitem{Louis:2002ny}
  J.~Louis and A.~Micu,
  ``Type II theories compactified on Calabi--Yau threefolds in the presence  of
  background fluxes,''
  Nucl.\ Phys.\  B {\bf 635} (2002) 395
  [\arXiv{hep-th/0202168}].

\bibitem{KPM}
   A.K.~Kashani-Poor and R.~Minasian,
   ``Towards reduction of type II theories on $\SU(3)$ structure manifolds,''
   JHEP {\bf 0703} (2007) 109
  [\arXiv{hep-th/0611106}].

\bibitem{DAuria}
   R.~D'Auria, S.~Ferrara and M.~Trigiante,
   ``On the supergravity formulation of mirror symmetry in generalized
   Calabi--Yau manifolds,''
   Nucl.\ Phys.\  B {\bf 780} (2007) 28
   [\arXiv{hep-th/0701247}].

\bibitem{MK}
   P.~Koerber and L.~Martucci,
   ``From ten to four and back again: how to generalize the geometry,''
   JHEP {\bf 0708} (2007) 059
   [\arXiv{0707.1038} [hep-th]].

\bibitem{BC}
   D.~Cassani and A.~Bilal,
   ``Effective actions and $N=1$ vacuum conditions from $\SU(3)\times\SU(3)$
   compactifications,''
   JHEP {\bf 0709}, 076 (2007)
   [\arXiv{0707.3125} [hep-th]], 

   D.~Cassani,
   ``Reducing democratic type II supergravity on $\SU(3)\times\SU(3)$
   structures,'' 
   JHEP {\bf 0806} (2008) 027
   [\arXiv{0804.0595} [hep-th]].

\bibitem{CKP}
  D.~Cassani and A.~-K.~Kashani-Poor,
  ``Exploiting N=2 in consistent coset reductions of type IIA,''
  Nucl.\ Phys.\ B {\bf 817} (2009) 25
  [\arXiv{0901.4251} [hep-th]].

\bibitem{CdO}
   A.~Strominger,
   ``Yukawa Couplings In Superstring Compactification,''
   Phys.\ Rev.\ Lett.\  {\bf 55} (1985) 2547, 

   A.~Strominger,
   ``Special Geometry,''
   Commun.\ Math.\ Phys.\  {\bf 133} (1990) 163, 

   P.~Candelas and X.~de la Ossa,
   ``Moduli Space Of Calabi--Yau Manifolds,''
   Nucl.\ Phys.\ B {\bf 355}, 455 (1991).

\bibitem{Strominger}
  A.~Strominger,
  ``Loop corrections to the universal hypermultiplet,''
  Phys.\ Lett.\  B {\bf 421} (1998) 139
  [\arXiv{hep-th/9706195}].

\bibitem{Antoniadis:1997eg}
  I.~Antoniadis, S.~Ferrara, R.~Minasian and K.S.~Narain,
  ``$R^4$ couplings in M- and type II theories on Calabi--Yau spaces,''
  Nucl.\ Phys.\  B {\bf 507} (1997) 571
  [\arXiv{hep-th/9707013}].

\bibitem{Gunther:1998sc}
  H.~G\"unther, C.~Herrmann and J.~Louis,
  ``Quantum corrections in the hypermultiplet moduli space,''
  Fortsch.\ Phys.\  {\bf 48} (2000) 119
  [\arXiv{hep-th/9901137}].

\bibitem{Becker:2002nn}
  K.~Becker, M.~Becker, M.~Haack and J.~Louis,
  ``Supersymmetry breaking and alpha-prime corrections to flux induced potentials,''
  JHEP {\bf 0206} (2002) 060
  [\arXiv{hep-th/0204254}].

\bibitem{AMTV}
  I.~Antoniadis, R.~Minasian, S.~Theisen and P.~Vanhove,
  ``String loop corrections to the universal hypermultiplet,''
  Class.\ Quant.\ Grav.\  {\bf 20} (2003) 5079
  [\arXiv{hep-th/0307268}].

\bibitem{loop}
  D.~Robles-Llana, F.~Saueressig and S.~Vandoren,
  ``String loop corrected hypermultiplet moduli spaces,''
  JHEP {\bf 0603}, 081 (2006)
  [\arXiv{hep-th/0602164}].

\bibitem{Andrianopoli:1996cm}
  L.~Andrianopoli, M.~Bertolini, A.~Ceresole, R.~D'Auria, S.~Ferrara,
  P.~Fr\'e and T.~Magri, 
  ``$N = 2$ supergravity and $N = 2$ super Yang-Mills theory on
  general scalar manifolds: Symplectic covariance, gaugings and the
  momentum map,'' 
  J.\ Geom.\ Phys.\  {\bf 23} (1997) 111
  [\arXiv{hep-th/9605032}].

\bibitem{deWit:1984pk}
   B.~de~Wit and A.~Van~Proeyen, 
   ``Potentials and symmetries of general
   gauged $N=2$ supergravity: Yang-Mills models,'' 
   Nucl.\ Phys.\ {\bf B245} (1984)  89.

\bibitem{Craps:1997gp}
  B.~Craps, F.~Roose, W.~Troost and A.~Van Proeyen,
  ``What is special Kaehler geometry?''
  Nucl.\ Phys.\  B {\bf 503}, 565 (1997)
  [\arXiv{hep-th/9703082}].

\bibitem{Bagger:1983tt}
   J.~Bagger and E.~Witten, ``Matter couplings in $N=2$ supergravity,''  
   Nucl.\ Phys.\ {\bf B222} (1983) 1.

\bibitem{deWit:1984px}
   B.~de~Wit, P.~G. Lauwers and A.~Van~Proeyen, 
   ``Lagrangians of $N=2$  supergravity -- matter systems,''
   Nucl.\ Phys.\ {\bf B255} (1985) 569.

\bibitem{CFG}
   S.~Cecotti, S.~Ferrara and L.~Girardello,
   ``Geometry of Type II Superstrings and the Moduli of Superconformal
   Field Theories,'' 
   Int.\ J.\ Mod.\ Phys.\  A {\bf 4}, 2475 (1989).

\bibitem{FS}
   S.~Ferrara and S.~Sabharwal,
   ``Dimensional Reduction Of Type II Superstrings,''
   Class.\ Quant.\ Grav.\  {\bf 6} (1989) L77, 

   S.~Ferrara and S.~Sabharwal,
   ``Quaternionic Manifolds For Type II Superstring Vacua Of
   Calabi--Yau Spaces,'' 
   Nucl.\ Phys.\ B {\bf 332} (1990) 317.

\bibitem{Strominger:1995cz}
  A.~Strominger,
  ``Massless black holes and conifolds in string theory,''
  Nucl.\ Phys.\ B {\bf 451} (1995) 96
  [\arXiv{hep-th/9504090}].

\bibitem{GGV}
M.B.~Green, M.~Gutperle and P.~Vanhove, ``One loop in eleven dimensions,'' 
Phys.\ Lett.~\textbf{B409} (1997) 177  [\arXiv{hep-th/9706175}].

\bibitem{Liu:2013dna} 
  J.T.~Liu and R.~Minasian,
  ``Higher-derivative couplings in string theory: dualities and the
  $B$-field,'' 
  [\arXiv{1304.3137} [hep-th]].

\bibitem{gemmer}
   K.-P. Gemmer, 
   ``Orientifolds and $R^4$-couplings on generalized geometries,'' 
   Hamburg University diploma thesis (2010),
   \href{http://www.desy.de/uni-th/stringth/Works/Gemmer\_Diplomarbeit.pdf}%
{www.desy.de/uni-th/stringth/Works/Gemmer\_Diplomarbeit.pdf}.

\bibitem{KashaniPoor:2013en}
  A.K.~Kashani-Poor, R.~Minasian and H.~Triendl,
  ``Enhanced supersymmetry from vanishing Euler number,''
  JHEP {\bf 1304} (2013) 058
  [\arXiv{1301.5031} [hep-th]].

\bibitem{KashaniPoor:2005si}
  A.K.~Kashani-Poor and A.~Tomasiello,
  ``A stringy test of flux-induced isometry gauging,''
  Nucl.\ Phys.\  B {\bf 728} (2005) 135
  [\arXiv{hep-th/0505208}].

\bibitem{HKC}
   B.~de Wit, B.~Kleijn and S.~Vandoren,
  ``Superconformal hypermultiplets,''
  Nucl.\ Phys.\  B {\bf 568} (2000) 475
  [\arXiv{hep-th/9909228}], 

  B.~de Wit, M.~Rocek and S.~Vandoren,
  ``Hypermultiplets, hyperk\"ahler cones and quaternion-K\"ahler
  geometry,'' 
  JHEP {\bf 0102}, 039 (2001)
  [\arXiv{hep-th/0101161}], 

  B.~de Wit, M.~Rocek and S.~Vandoren,
  ``Gauging isometries on hyperK\"ahler cones and quaternion-K\"ahler
  manifolds,''
  Phys.\ Lett.\  B {\bf 511}, 302 (2001)
  [\arXiv{hep-th/0104215}].

\bibitem{Bohm:1999uk}
  R.~B\"ohm, H.~G\"unther, C.~Herrmann and J.~Louis,
  ``Compactification of type IIB string theory on Calabi--Yau threefolds,''
  Nucl.\ Phys.\  B {\bf 569} (2000) 229
  [\arXiv{hep-th/9908007}].

\bibitem{TV}
U.~Theis and S.~Vandoren,
  ``$N=2$ supersymmetric scalar tensor couplings,''
  JHEP {\bf 0304}, 042 (2003)
  [\arXiv{hep-th/0303048}].

\end{thebibliography}
\end{document}